\PassOptionsToPackage{dvipsnames}{xcolor}

\newif\ifarxiv
\arxivtrue

\documentclass[conference]{IEEEtran}
\ifarxiv\IEEEoverridecommandlockouts\fi %

\usepackage{amsmath}
\usepackage{amssymb}
\usepackage{algorithm}
\usepackage{algorithmic}
\usepackage{graphicx}
\usepackage{textcomp}
\usepackage{xcolor}
\usepackage[all]{nowidow}
\usepackage{mathtools}
\usepackage{bm}
\usepackage{fnpct} %
\usepackage{pifont}
\usepackage{booktabs}
\usepackage{siunitx}

\usepackage[numbers,sort&compress]{natbib}

\usepackage[hidelinks]{hyperref}
\usepackage[capitalise,noabbrev]{cleveref}

\usepackage{makecell}
\usepackage[flushleft]{threeparttable}
\usepackage{multirow}
\usepackage{tabularx}
\usepackage{tikz}
\usetikzlibrary{patterns.meta}

\newcolumntype{Y}{>{\centering\arraybackslash}X}

\usepackage{xspace}
\newcommand{\sysnameplain}{AdapBench}

\newcommand{\sysname}{\sysnameplain\xspace}

\begin{document}

\title{\sysname: Adaptive Early-Stopping and Prioritization for Efficient Microbenchmarking in Cloud-based CI/CD}

\author{%
\IEEEauthorblockN{Sebastian Koch, Jonas Heisterberg, Nils Japke, Trever Schirmer, and David Bermbach}
\IEEEauthorblockA{\textit{Scalable Software Systems Research Group}, \textit{TU Berlin}, Berlin, Germany \\
\{sk, jh, nj, ts, db\}@3s.tu-berlin.de}
\ifarxiv\thanks{\copyright{} 2026 IEEE. Personal use of this material is permitted. Permission from IEEE must be obtained for all other uses, in any current or future media, including reprinting/republishing this material for advertising or promotional purposes, creating new collective works, for resale or redistribution to servers or lists, or reuse of any copyrighted component of this work in other works. Accepted for publication at the 19th IEEE/ACM International Conference on Utility and Cloud Computing (UCC 2026).}\fi
}

\maketitle

\begin{abstract}
Cloud-based software systems frequently release new versions, a fact further accelerated through agentic coding.
Each new release, however, brings the risk of performance regressions which can irritate end users, increase infrastructure costs, and increase carbon footprints.
Therefore, developers strive to catch such regressions in testing stages, usually through so-called continuous benchmarking.

The standard way of doing this is using microbenchmark suites which, however, take notoriously long to execute.
Optimizing the execution of microbenchmark suites is, hence, an active field of research.

In this paper, we propose AdapBench, an online scheduling system for reducing the execution duration of microbenchmark suites.
AdapBench is the first approach which dynamically makes at-runtime decisions for prioritizing the most informative benchmarks and early-stopping each one once enough evidence has accumulated to accept or reject a performance change hypothesis.

We evaluate our prototype on eight popular Java projects, commonly used by microbenchmark research, and four additional Go projects.
At matched \qty[mode=match, reset-text-series=false, text-series-to-math=true]{90}{\percent} change detections and false-positive rate compared to an optimistic uniform run, \sysname{} saves \qty[mode=match, reset-text-series=false, text-series-to-math=true]{42}{\percent} of the aggregate execution duration on the Java projects, with negligible scheduling overhead (below \qty[mode=match, reset-text-series=false, text-series-to-math=true]{0.001}{\percent} of the total runtime).
On the Go projects, \sysname{} finds every performance change with zero false positives while saving \qty[mode=match, reset-text-series=false, text-series-to-math=true]{85}{\percent} of the full run.
\end{abstract}


\hypersetup{pdftitle={AdapBench: Adaptive Early-Stopping and Prioritization for Efficient Microbenchmarking in Cloud-based CI/CD}}

\section{Introduction}
\label{sec:introduction}

\begin{figure}
  \centering
  \includegraphics[width=0.78\columnwidth]{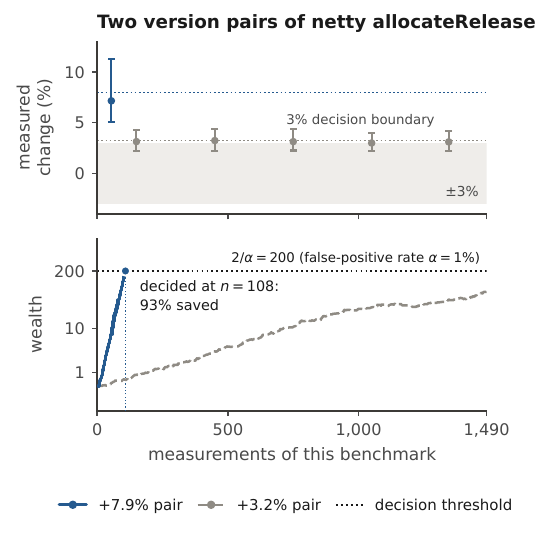}
  \caption{We accumulate evidence for two benchmarks answering whether a performance change is present. Since one benchmark exhibits a bigger performance change (\qty{+7.9}{\percent}), the evidence accumulates faster and crosses the decision threshold after $108$ of $1490$ measurements. For the second benchmark (\qty{+3.2}{\percent}), it moves around the \qty{3}{\percent} decision boundary, so the evidence accumulates slowly and the benchmark cannot be early-stopped.}
  \label{fig:teaser}
\end{figure}
With the growing adoption of AI-assisted development, code changes to cloud services ship more frequently than ever~\cite{daniotti2026aicode}.
Each change can introduce performance regressions that functional tests miss.
Consequently, performance assurance must become more frequent and more rigorous.
Without a performance assurance gate, regressions will go increasingly unnoticed, driving up operational costs.
Especially in large-scale cloud services, performance drift compounds quickly.

One method of assuring performance is microbenchmarks, which measure the performance of small code fragments such as individual methods~\cite{LaaberMBEval}.
Due to their small testing surface, microbenchmarks enable better root cause analysis than benchmarks of the whole application.
Unlike unit tests that have a binary outcome after a single repetition, microbenchmarks need to be executed many times to yield a statistically meaningful result: a single run of a microbenchmarking suite can take hours to execute~\cite{LaaberMBEval}.
Accordingly, reducing the time to execute such suites is an active area of research~\cite{laaber2020dynamically}.
While the current model employed by practitioners is to run all benchmarks uniformly for a fixed duration~\cite{leitner2017exploratory, daly_industry_2019}, recent research has used historical information to reduce the execution time of microbenchmarks offline by identifying when a benchmark is statistically stable~\cite{japke2025muoptimestaticallyreducingexecution}.
However, their approach requires a relatively stable codebase such that historical information serves as meaningful priors.
It would be ideal to know, a priori, how many measurements are required per microbenchmark.
This is, however, infeasible in practice as execution environments are noisy and unpredictable, especially in the cloud~\cite{leitner2016patternschaosstudy, Laaber_2019}.

As a solution, we propose \sysname{}, a complementary approach that solely relies on dynamic runtime information, independent of historical data or other input features such as code changes to decide on the order of execution and to early-stop microbenchmark execution when a change is statistically proven.
As such, our approach is well-equipped for the dynamic performance characteristics present in noisy cloud VMs where CI pipelines typically run~\cite{leitner2016patternschaosstudy, Laaber_2019}.

We optimize microbenchmark suite execution along two axes: online benchmark prioritization and execution depth via early stopping.
Some benchmarks are more informative than others. 
They accumulate evidence about a performance regression faster and can therefore be stopped after fewer samples, as \cref{fig:teaser} illustrates.
Their execution is prioritized to uncover regressions sooner.
As a result, we identify performance regressions more quickly and cost-effectively.

The challenge is to adaptively identify those informative benchmarks at runtime.
We address this with e-processes~\cite{ramdas2023gametheoretic}, with which we construct an anytime-valid confidence sequence that yields a reliable per-benchmark stopping signal without running the benchmark to exhaustion.
In summary, we make the following contributions:%
\begin{itemize}
  \item We frame microbenchmark suite execution as an online optimization problem in which we leverage anytime-valid e-processes to support per-benchmark early stopping and online prioritization.
  \item We implement and open-source a system that optimizes suites with pluggable prioritization policies (\cref{sec:system-design}).
  \item We perform extensive experimentation, empirically validating a decision-closest prioritization policy that surfaces performance changes $1.7\times$--$4.3\times$ faster than round-robin scheduling on Java and up to $5.9\times$ using Go. Our scheduler holds the false-positive rate 
    and saves \qty{42}{\percent} of the aggregate measurement budget at matched recall against the best uniform configuration for Java (\qtyrange{27}{61}{\percent} per project); on Go, it saves \qty{85}{\percent} of the full run at full recall with zero false positives (\cref{sec:evaluation}).
\end{itemize}

\section{Background and Related Work}
We highlight both background and current research directions in the following section.

\label{sec:background}
\subsection{Cloud Service Benchmarking}
One tool to assess the quality of cloud services is benchmarking, in which systems are evaluated under stress scenarios~\cite{Bermbach_book_2017}.
The field distinguishes between two types of cloud service benchmarks: microbenchmarks and application benchmarks.
Application benchmarks evaluate system qualities at external interfaces by stressing the system through realistic end-to-end scenarios that resemble production use.
Central to this work, microbenchmarks repeatedly execute small, isolated pieces of software, enabling the quality evaluation for a subset of features~\cite{LaaberMBEval}.
In contrast to application benchmarks, microbenchmarks yield finer-grained measurements, and therefore a full suite execution covering the complete code base is costly and time-consuming.
The targeted qualities might differ based on system requirements~\cite{Bermbach_book_2017}; however, in this paper, we focus on latency as a performance metric.

Performance benchmarking is an important step in the life cycle of a cloud service, as hidden performance regressions can compound into high operational costs at scale or degrade user experience.
As part of code review, common practice involves executing microbenchmarks in a pull request's CI pipeline to identify significant performance deviations.
Typically, the CI pipeline spins up a virtual machine and runs the microbenchmark for a fixed-$n$ duration decided a priori~\cite{daly_industry_2019, javed_perfci_2020}.

If practitioners run their microbenchmarks on a virtual machine, influencing factors such as noisy neighbors make it impractical to evaluate the cloud service performance in absolute terms unless those factors are accounted for.
A prominent approach to address this issue is duet benchmarking~\cite{Bulej_2020_Duet_Benchmarking}.
Instead of comparing absolute performance measurements, we calculate the difference and quantify performance changes in relative terms.
Duet benchmarking executes the benchmark in the same environment with synchronized workloads.
Beyond spatial noise, cloud benchmarks also exhibit strong temporal dependence, which manifests as autocorrelation in the measurements~\cite{Schirmer_2023, leitner2016patternschaosstudy}.
Repeatedly executing the same benchmark in sequence therefore does not resemble a real-world scenario, as caching effects, for instance, speed up executions over time.
Randomized multiple interleaved trials (RMIT)~\cite{RMIT} mitigate this by randomly shuffling the execution order of different benchmarks instead of running them sequentially.

\subsection{Performance Change Detection}
\label{subsubsec:change-detection}
A common use of benchmarks is to detect whether a system's performance has changed between two software versions.
To this end, multiple statistical approaches exist to compare two sets of measurements and report whether there is a significant change in a specific statistic.
These statistics are typically estimators, such as the mean or median, that indicate performance tendencies in a system.

Especially in cloud services, performance measurements do not follow a specific distribution, and therefore non-parametric methods are favorable.
To address this, bootstrapped confidence intervals are an established method~\cite{Davison_Hinkley_1997} that resamples the original measurements with replacement to calculate a confidence interval for the statistic of interest.
If the resulting confidence interval excludes a defined change band, a significant change is reported.
Following previous research, we define a significant performance change as any deviation beyond $\pm \qty{3}{\percent}$~\cite{japke2025muoptimestaticallyreducingexecution}.

\subsection{Anytime Valid Inference and Early Stopping}
State-of-the-art benchmarking often finds a predefined duration for how long a benchmark should run and then runs it to exhaustion~\cite{georges2007statistically}.
The field proposed methods to approximate the ideal duration before running an experiment~\cite{japke2025muoptimestaticallyreducingexecution}.
These, however, do not dynamically adjust to changes in the execution environment and require historical information.
Therefore, online adaptation serves as a powerful, complementary method.
What makes online methods challenging is that the false-positive guarantee of classical statistical methods collapses under peeking~\cite{johari2022always}.
A performance change detection system in practice is usually only useful when it operates under low false-positive rates.
To establish an online optimization, we use sequential testing with betting confidence sequences~\cite{wald1945sequential, shafer2021testing, waudbysmith2024estimating}.
This method is anytime-valid, meaning we can peek at the data at any point in time and still make a decision with false-positive control~\cite{ramdas2023gametheoretic, howard2021timeuniform}.

With betting confidence sequences we grow a wealth process~\cite{shafer2021testing, waudbysmith2024estimating}.
Formally, an e-value is a non-negative measure of evidence against the null hypothesis whose expectation under the null is at most one.
The wealth process through betting is exactly a sequence of such e-values, an e-process that stays valid at every point in time.
By Ville's inequality, rejecting the first time the e-process $\ge 1/\alpha$ controls the false-positive rate at level $\alpha$ no matter when we stop.
We bet across a grid of possible mean values simultaneously; the set of means not yet rejected forms an anytime-valid \emph{confidence sequence}.
If the confidence sequence lies beyond the \qty{3}{\percent} band, we report a change.

\subsection{Related Work}
\label{sec:related-work}

\paragraph*{Software Microbenchmarking and Measurement Rigor}
Software microbenchmarking is a well-established method to assure software performance.
Laaber et al.~\cite{Laaber_2019} present the challenges of performance variability when running microbenchmarks in the cloud.
Microbenchmarks need many repetitions to account for the changing performance characteristics of the underlying infrastructure.
\citet{Schirmer_2023} show performance differences of up to \qty{15}{\percent} for the same workload on Google Cloud Functions due to temporal patterns, measured over a corpus spanning several months.
Abedi and Brecht~\cite{RMIT} suggest RMIT to mitigate some of those issues, which we use as part of our experiment design.
Kalibera and Jones~\cite{kalibera2020quantifying} highlight the complementary statistical toolkit.
They flag the misuse of confidence intervals and ignored variance among other problems, and recommend bootstrap CIs.

\paragraph*{Cloud Measurement and Performance Variability}
Duet Benchmarking~\cite{Bulej_2020_Duet_Benchmarking}, as already described, mitigates cloud variability by running two software versions in the same execution environment under workload synchronization.
In the same vein, Japke et al.~\cite{japke2025isolation} evaluate the impact of different isolation strategies under duet benchmarking.
In practice, our system natively hooks into isolated, synchronized RMIT duet benchmarks.

\paragraph*{Reducing Benchmarking Cost}
Prior work on reducing the execution time of microbenchmark suites~\cite{laaber2020dynamically, japke2025muoptimestaticallyreducingexecution} underpins the relevance for our approach.
Laaber et al.~\cite{laaber2020dynamically} early-stop a benchmark online once its measurements are statistically stable; for instance, when the relative confidence interval width (RCIW) converges within a sliding window and further runs yield little new information.
Alghmadi et al.~\cite{alghmadi2016stop} track repetitiveness over long-horizon performance tests and decide when to stop.
Similar to other prior work, this approach has no inherent error control baked in and targets long performance tests, not microbenchmark suites.
Inspired by test case prioritization (TCP), Laaber et al.~\cite{laaber2024prioritization} rerank the execution order, prioritizing the benchmarks that surface regressions sooner.
Prado Lima and Vergilio~\cite{pradolima2020coleman} model test case prioritization as a multi-armed bandit problem, which is similar to our modeling.
They track which tests fail often and order them accordingly.
However, functional tests execute once, have no depth, no measurement distribution, and need no error control.

\paragraph*{Performance Regression Detection and Continuous Benchmarking}
The industry uses benchmarks in CI pipelines to detect regressions~\cite{daly_industry_2019, Japke_2023}.
CI makes continuous performance assurance feasible, particularly in fast-changing software~\cite{continuous_benchmarking_grambow, javed_perfci_2020, schirmer2024elastibench, rese2024increasingefficiencyresultreliability, japke2025cicd}.
However, current practice involves fixed-$n$ benchmarks with no adaptive online scheduling loop.

\paragraph*{Anytime-Valid Inference and E-Values}
Wald's sequential probability ratio test~\cite{wald1945sequential} introduces the notion of sequential hypothesis testing.
Modern anytime-valid inference generalizes the idea by accumulating evidence as e-values, which in turn yield confidence sequences~\cite{howard2021timeuniform, ramdas2023gametheoretic}.
We adopt betting confidence sequences~\cite{shafer2021testing, waudbysmith2024estimating} for our scheduler, keeping the underlying statistics unchanged.
Anytime-valid statistics are already commonly used in A/B testing and monitoring~\cite{johari2022always}, but no prior work has adopted e-values for cloud service benchmarking.

\section{System Design: \sysnameplain}
\label{sec:system-design}
The primary target of our approach is evolving software, where frequent code changes demand continuous performance assurance.
We therefore measure relative performance changes between two versions by sampling paired measurements and computing their difference~\cite{Bulej_2020_Duet_Benchmarking}.
Our objective is to match the false-positive and false-negative rates of a full run while spending substantially less execution time.

To this end, our approach adapts anytime-valid e-processes to make informed online scheduling decisions under strict error control.
Each measurement provides evidence against the null hypothesis that a specific benchmark shows no relevant performance change.
These statistically valid online decisions let us optimize along two axes, namely early stopping and prioritization.
Our method requires no prior project knowledge or historical data.
Instead, it relies purely on exploration during the iterative benchmark execution and learns online which benchmarks reveal performance regressions.

We propose a scheduler that runs benchmarks iteratively and at each step greedily selects the one nearest a decision, so that each new measurement informs the next scheduling decision in a closed feedback loop.
We evaluate the scheduler's overhead in \cref{subsec:overhead}.
\Cref{alg:scheduler} summarizes the loop; the following subsections detail its steps.
The section follows one pass through this loop: \cref{subsec:evidence} details
how each measurement becomes evidence, \cref{subsec:stopping} when a benchmark
stops, \cref{subsec:prioritization} which benchmark runs next,
\cref{subsec:error-control} why false-positive control holds, and
\cref{subsec:implementation} how we implement the scheduler.

\begin{algorithm}
\caption{\sysname online scheduling loop }
\label{alg:scheduler}
\begin{algorithmic}[1]
\REQUIRE benchmarks $\mathcal{A}$ with per-benchmark budgets $B_a$, level $\alpha$, change band $\varepsilon$
\ENSURE per-benchmark decision (changed / no change / undecided)
\STATE wealth $K_a(m) \gets 1$ for $m = \pm\varepsilon$,\; $\mathrm{active}_a \gets \mathrm{true}$ for all $a \in \mathcal{A}$
\WHILE{some benchmark is active}
  \STATE $a^\ast \gets$ active benchmark nearest a decision \COMMENT{occasional fewest-samples exploration}
  \STATE $X_t \gets \textsc{Measure}(a^\ast)$ \COMMENT{mean of one block of duet differences (\cref{eq:duet})}
  \STATE $K_{a^\ast}(m) \gets K_{a^\ast}(m)\bigl(1 + \lambda_t(m)\,(X_t - m)\bigr)$ for $m = \pm\varepsilon$ \COMMENT{\cref{eq:agrapa,eq:agrapa-bet}}
  \IF{confidence interval excludes $[-\varepsilon, +\varepsilon]$}
    \STATE retire $a^\ast$: changed \COMMENT{claim}
  \ELSIF{confidence interval contained in $(-\varepsilon, +\varepsilon)$}
    \STATE retire $a^\ast$: no change \COMMENT{futility}
  \ELSIF{no claim reachable within $B_{a^\ast}$}
    \STATE retire $a^\ast$: undecided \COMMENT{horizon stop}
  \ELSIF{$a^\ast$ has spent $B_{a^\ast}$}
    \STATE retire $a^\ast$: fixed-$n$ bootstrap CI decides \COMMENT{fallback}
  \ENDIF
\ENDWHILE
\end{algorithmic}
\end{algorithm}

\subsection{Evidence Accumulation per Benchmark}
\label{subsec:evidence}
Formally, for paired measurements $s_1,s_2$ of the two versions we record the \emph{duet difference}
\begin{equation}
  d = \frac{s_1 - s_2}{s_1 + s_2} \in (-1, 1),
  \qquad
  s_1 = (1+r)\,s_2 \;\Leftrightarrow\; d = \frac{r}{2+r}.
  \label{eq:duet}
\end{equation}
A benchmark counts as changed when its true mean $\mu = \mathbb{E}[d]$ leaves the change band, $|\mu| > \varepsilon$; the \qty{3}{\percent} threshold thus maps to $\varepsilon = 0.03/2.03 \approx 0.015$, and we report magnitudes on the ratio scale via the inverse $r = 2d/(1-d)$.
Our null hypothesis is that there is no performance change greater than \qty{3}{\percent}.
To establish an anytime-valid confidence sequence, we represent each benchmark as a wealth process.
If the null hypothesis holds, the process behaves like a fair game: as we bet against the null, we neither gain nor lose wealth on average.
If the benchmark uncovers a performance change, we win more often than we lose, and our wealth accumulates.
Once our wealth reaches a threshold, we can confidently report the benchmark as a significant change in either direction.
Alternatively, another relevant null hypothesis in practice is that there is no performance regression greater than \qty{3}{\percent} while ignoring performance improvements.

The betting process is valid if the bet is only dependent on past observations, i.e., we cannot place a bet considering the current measurement.
Under suboptimal betting the evidence accumulates too slowly, which means our approach does not translate into cost savings, i.e., it takes too long to decide.
Our modular architecture supports multiple betting strategies through adapters, but we opt for a growth-optimal strategy (Kelly betting).
However, Kelly betting assumes we know the probabilities of winning.
Since we learn the outcomes over time, we opt for an approximate Growth Rate Adaptive to the Particular Alternative (aGRAPA) rule~\cite{waudbysmith2024estimating} that learns them online.
We break down our strategy: if a change appears to be big and the measurements show little variance, we bet big.
If, on the other hand, the measurements have high variance and the change is small, we bet little.
Using this strategy, we can uncover big performance changes quickly, while for borderline changes we might not save measurements by design.

For a candidate mean change $m$ and an observation $X_t$ the wealth multiplies by $K_t(m)$ defined as:
\begin{equation}
  K_t(m) = K_{t-1}(m)\bigl(1 + \lambda_t(m)\,(X_t-m)\bigr).
  \label{eq:agrapa}
\end{equation}
The initial wealth starts at $K_0(m)=1$.
The bet size $\lambda_t(m)$ follows aGRAPA:
\begin{equation}
  \lambda_t(m) \;=\; \frac{\hat{\mu}_{t-1} - m}{\hat{\sigma}^{2}_{t-1} + (\hat{\mu}_{t-1} - m)^{2}},
  \label{eq:agrapa-bet}
\end{equation}
where $\hat{\mu}_{t-1}$ and $\hat{\sigma}^{2}_{t-1}$ are the running mean and variance of the past observations $X_1,\dots,X_{t-1}$.
The bet grows with the estimated mean change and shrinks with variance.
We clip each duet-difference observation (\cref{eq:duet}) to $\pm 0.06$ as the betting process requires bounded observations~\cite{waudbysmith2024estimating}.
The confidence sequence therefore does not estimate the true effect size beyond $\pm 0.06$.

We average raw measurements as a block of size $b$ and feed one block mean as the $X_t$ observation.
In doing so, we lower the variance and improve the betting behavior.
We additionally reduce autocorrelation of the raw measurements.
Depending on the experiment setup and the noisiness of the expected execution environment, we can increase or decrease the block size.

Following \cref{eq:agrapa-bet}, each state update requires the running mean and variance and runs in $O(1)$ time.

\subsection{Early Stopping and Decision}
\label{subsec:stopping}
To effectively save execution time, we early-stop benchmarks that reach a verdict before the budget is exhausted.

To this end, we hold a set of mean candidates and test for each candidate the null that the true mean is $m$.
As we sample from the benchmarks, we can reject some of the mean candidates; the surviving ones form an anytime-valid confidence interval.
Rejection happens when the betting wealth (evidence) $K_t(m)$ exceeds $\tau = 2/\alpha$~\cite{waudbysmith2024estimating}.
Instead of tracking all possible candidate means, our algorithm (\cref{alg:scheduler}) watches the two edges at $\pm\varepsilon$ to be more efficient.
We apply the performance change detection logic from \cref{subsubsec:change-detection}.
If the confidence interval excludes the $\pm \qty{3}{\percent}$ band, we report a change.
Using the benchmark's confidence interval and the change band $\varepsilon$ ($\varepsilon \approx 0.015$, a \qty{3}{\percent} change on the ratio scale, \cref{eq:duet}) we design the following three stopping rules:
(1)~\emph{Claim} --- The confidence interval excludes $\pm \qty{3}{\percent}$ and we claim a change.
(2)~\emph{Futility} --- The confidence interval narrows to within $\pm \qty{3}{\percent}$ and we report no change.
(3)~\emph{Horizon} --- We expect no decision to be reached within the budget, so we retire the benchmark as undecided.

As we claim changes as part of an anytime-valid e-process, no predefined sample size is required and we can adapt to the data and early-stop at any point with false-positive control.

If we cannot reach a decision before the full run is exhausted, we use a regular bootstrap confidence interval, just as practitioners do today.
The anytime-valid confidence sequence pays a tax and is generally less powerful than a regular test; therefore after a full run we default to the baseline approach to not lose power.
We defer controlling the two resulting $\alpha$ levels to \cref{subsec:error-control}.

\paragraph{Horizon Stopping}
Intuitively, when a benchmark cannot reach a decision within its budget, running it further is hopeless, and it is reasonable to stop it.
We take the running mean and add $z$ (between $1$ and $1.64$ adapting to the measurement count) standard errors, an optimistic effect estimate that the data still supports.
We then calculate the gap between this estimate and the \qty{3}{\percent} decision boundary.
Assuming near-optimal wealth (evidence) growth, if the benchmark cannot reach the claim threshold within its full horizon at two consecutive checks, we retire the benchmark as undecided.
Horizon stopping only retires benchmarks and never claims them.
This has potentially negative consequences for recall, but never for the controlled false-positive rate.
To limit recall loss, we place the first checkpoint at \qty{50}{\percent} of the budget, a conservative configuration.

\subsection{Decision-Closest Prioritization}
\label{subsec:prioritization}
Beyond early stopping, we optimize the benchmark execution prioritization with the aim to surface performance regressions sooner.
As an intuition: if we compare two benchmarks where one shows a \qty{30}{\percent} regression in the first few measurements with low variance and the other stays around \qty{0}{\percent}, we want to first take a closer look at the \qty{30}{\percent} one.

A side effect of early stopping is that we occasionally stop the wrong benchmarks, especially under the horizon stopping. %
Once a benchmark is stopped, we cannot recover its measurements, which risks missing regressions.
Prioritization compensates on a complementary axis, surfacing regressions sooner, which helps absorb the recall loss from wrong stops.

Selecting the next benchmark to run, however, presents a trade-off between exploration and exploitation.
Without historical data, we first need to explore which benchmarks are more useful than others.
Then, we can exploit this knowledge by running the most informative benchmarks first.

For each execution decision, \sysname scores each benchmark and selects the lowest score as the next execution step.
We calculate the score by estimating the number of measurement blocks until we expect the benchmark to reach a verdict.
If a benchmark is already within the $\pm \qty{3}{\percent}$ band, it gets pushed towards the end of the execution.
We apply the same projection as for the horizon stopping to estimate the blocks, here with the plain running mean instead of the optimistic estimate.
To account for the exploration/exploitation trade-off we assign \qty{15}{\percent} of steps to pure exploration.
The exploration step selects the benchmark with the fewest measurements so far, so no benchmark is starved.
Since the projection requires a variance estimate which is unreliable at low measurement counts, we take the biggest observed change for the cold start of the experiment (under 3 blocks).
The scoring reads the running mean and variance but never the current evidence (wealth), such that the e-process stays valid.
Once enough evidence has accumulated, early stopping engages and frees up budget.

Our system is agnostic to the specific policy, and others might incorporate historical project data or baseline reductions; we deliberately choose a simple, fast policy first and leave alternatives to future work.

\subsection{Error Control}
\label{subsec:error-control}
One of the strengths of our approach is the false-positive control we gain by employing e-processes.
In contrast, prior work~\cite{laaber2020dynamically, japke2025muoptimestaticallyreducingexecution} often optimizes execution time without inherent error control baked into the system.
We derive decisions by looking at the verdict, not purely whether results are stable.
A verdict can be reached by early confidence sequence claims or through the regular bootstrap CI detection for undecided benchmarks.
Those mechanisms are mutually exclusive, i.e., an early-stopped benchmark will never be decided through the bootstrap CI mechanism.
As discussed earlier, we need to account for two $\alpha$ levels; we validate empirically in \cref{subsubsec:fpr-results} that the combined false-positive rate stays below $\alpha$.
Importantly, the false-positive bound is per benchmark and not on the whole suite.

Certain assumptions need to hold for our experiments to be valid, one of which is the independence of measurements, which RMIT addresses by reducing autocorrelation.
Since the scheduler decides the execution order, it can automatically enforce RMIT at invocation granularity.

\subsection{Implementation and Deployment}
\label{subsec:implementation}
A user specifies a budget for each benchmark in the suite, capping its number of measurements.
Within this budget, however, the scheduler decides after each measurement which benchmark to execute next.
The system holds a minimal state for each benchmark's evidence, making simple calculations for each scheduling decision.

The configuration for the scheduler includes the following:
(1)~the change band $\varepsilon$, i.e., at what point we consider a change relevant (in our case \qty{3}{\percent}, informed by prior work~\cite{japke2025muoptimestaticallyreducingexecution});
(2)~the false-positive rate $\alpha$ per benchmark (not suite level);
(3)~per-benchmark budget (an integer for the measurements);
(4)~block size to average over; and
(5)~whether we consider slowdowns only or changes in both directions.
Notably, new benchmarks can immediately benefit from the cost savings without any prior runs.

The scheduler acts on an abstraction and is independent of the actual underlying measurement engine.
Supporting a new language is accomplished with a new adapter;
the adapter simply implements a \texttt{measure} method that returns a duet measurement.

To increase efficiency, our scheduler holds two priority queues: one for choosing the benchmark with the lowest score and one for the least-measured one.
For each scheduling step, only the information about one benchmark changes, and we exploit this property.
Depending on whether we are in an exploitation or exploration step, we peek from the queue and take one measurement.
If we complete a block of measurements---depending on the block size---we calculate the new score and push to the queues.

Each scheduling step runs in $O(\log n)$ time with $n$ being the number of benchmarks.
When it comes to space-efficiency, we store the per-benchmark state. 
Additionally, we store each raw measurement since we do the bootstrap test at the end of the full exhaustive run for undecided benchmarks.
We can, however, drop the benchmarks' raw measurements for the ones we already decided.

\section{Evaluation}
\label{sec:evaluation}
To evaluate \sysname, we first implement a proof-of-concept (PoC) prototype and then conduct four experiments, each answering a specific research question.
Our open-source prototype\footnote{Prototype, experiment scripts, and the recorded Go dataset: \url{https://github.com/s3bastiankoch/adapbench}}, written in Python, replays the recorded benchmarks and implements the system design described in \cref{sec:system-design}.

We validate our approach along four research questions:
\begin{itemize}
  \item[\textbf{E1:}] Does \sysname operate under strict false-positive rate control?
  \item[\textbf{E2:}] How much execution time does \sysname save at matched recall and false positives against the best uniform baseline?
  \item[\textbf{E3:}] Does decision-closest prioritization surface regressions faster than round-robin scheduling?
  \item[\textbf{E4:}] Is the scheduling overhead small enough for practical use?
\end{itemize}

\subsection{Experiment Setup}
In our experiments, we use the prototype to replay pre-recorded benchmark data instead of running the benchmarks live.
This creates a controlled environment that ensures comparability between \sysname and the baseline, as both receive identical data.
Consequently, our results are unaffected by artifacts from run-to-run variation.

Throughout the evaluation, we use the following conventions:
A \emph{trace} is one CI pipeline run, i.e., one complete suite execution covering both software versions (A and B).
Each trace consists of multiple \emph{benchmark pairs}, where a pair comprises the same benchmark in both versions.
In turn, each benchmark is executed over many iterations within a fixed duration, and the mean over these iterations yields a single \emph{measurement}.

Finally, our experiments evaluate two regimes: the first considers all performance changes relevant; the second targets only regressions and not improvements, which is the more interesting scenario in practice.

\subsubsection{Datasets}
The microbenchmark datasets stem from two sources.
First, we use recorded microbenchmarks from eight Java projects established by prior work on benchmark test case prioritization (TCP)~\cite{laaber2021tcp}: bytebuddy, eclipsecollections, jctools, jenetics, netty, okio, xodus, and zipkin.
The Java benchmarks provide a median of \num{1490} measurements.
Second, we record our own dataset of four Go projects (prometheus/common, klauspost/compress, uber-go/zap, and go-toml) under RMIT to test our assumption of independent and identically distributed (i.i.d.) measurements.
For this dataset, we executed each benchmark pair for \num{500} measurements on GCP spot VMs.

\begin{figure}[t]
  \centering
  \includegraphics[width=0.78\columnwidth]{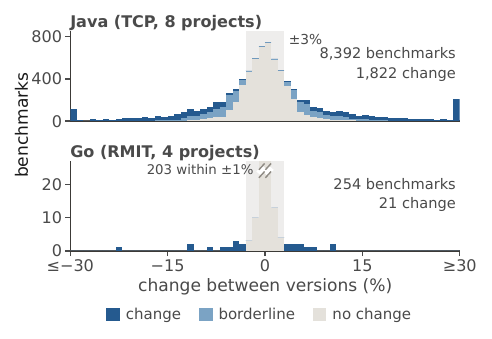}
  \caption{We highlight the distribution of changes and no-changes in the Java and Go datasets. Especially Go has many non-productive benchmarks which we can early-stop to save budget. Borderline marks benchmarks close to the \qty{3}{\percent} decision boundary that cannot be decided reliably.}
  \label{fig:dataset}
\end{figure}

\cref{fig:dataset} displays the evaluation corpus.
We derive change labels from exhaustive runs over all available measurements, which serve as the ground truth for evaluating the recall and false-positive rate of \sysname.
We define a relevant performance change as a mean change greater than \qty{3}{\percent}, a decision boundary informed by previous work~\cite{japke2025muoptimestaticallyreducingexecution}.
Specifically, we replay the full suite execution under resampling with replacement and perform a bootstrap CI test as described in \cref{sec:background}.
We label each version pair based on the resulting confidence interval.
We assign a \emph{change} if the CI lies entirely outside the \qty{3}{\percent} decision boundary and \emph{no change} otherwise.
The reference verdict replays the same measurement stream that \sysname consumes, so any disagreement traces to our early stopping rather than to sampling variation.
In \cref{fig:dataset}, \emph{borderline} marks benchmarks whose verdict flips across replays, i.e., benchmarks too close to the boundary to decide reliably.
For each project corpus, we exclude benchmarks with fewer than \num{60} measurements to ensure the reference labels rest on sufficient data.

\subsubsection{Replay model}
Prior work emphasizes RMIT~\cite{RMIT} to mitigate the effects of cloud variability and autocorrelation on benchmark results.
We therefore approximate RMIT in our experiments by using bootstrap resampling for the Java projects as they were recorded sequentially.
\sysname does not work on sequentially collected, autocorrelated measurements, as the anytime-valid confidence sequence construction requires i.i.d.\ data.
Since collection should follow RMIT, our i.i.d.\ assumption follows recommended practice.
Our self-collected dataset requires no resampling approximation since it directly used RMIT.
We simply replay it in the recorded order.

\subsubsection{Baseline}
\citet{laaber2020dynamically} and \textmu OpTime~\cite{japke2025muoptimestaticallyreducingexecution} optimize for measurement stability and expose no error-rate parameter, leaving no operating point at which to match recall and false positives.
\textmu OpTime furthermore requires historical data, excluded from our system by design.
We therefore omit both as baselines.
We also omit the prioritization approaches of \citet{laaber2024prioritization} and \citet{pradolima2020coleman}, as they require code-change features or failure history, which we exclude by design; our prioritization baseline is round-robin, as in today's CI pipelines.
Instead, we compare \sysname against current practice: a uniform configuration.
In practice, we cannot determine a priori how long each benchmark needs to run.
Therefore, we fall back to running all benchmarks sufficiently long to definitely surface the regressed benchmarks.
This fixed duration leaves savings on the table for the benchmarks that are easy to decide, which is what we exploit.
It is, however, infeasible to determine beforehand the easily decidable benchmarks, even with historical data, since code changes can fundamentally change the characteristics of a benchmark.
To establish a generous baseline, we identify the best uniform configuration ($N^\star$) that reaches a given recall level.
By recall level, we mean the fraction of true performance regressions we find.
To find the ideal $N^\star$ configuration we search a grid of different execution depths and select the best one, given a target recall level.
This baseline relies on an oracle and is thus overly optimistic compared to real CI environments, yet we consider it a fair baseline.
However, at shallow depth the bootstrap confidence intervals overclaim and inflate the false positives.
As a correction step, to ensure a fair baseline, we widen the baseline's CI to, for instance, \qty{99.3}{\percent}.
All comparisons are, therefore, made at matched false-positive rate.

\subsubsection{Experiments}

\paragraph{E1: False-Positive Control}
The experiment evaluates the question whether \sysname claims a performance change in the absence of an actual change below a rate $\alpha$ of \qty{1}{\percent}.
We construct an experiment that makes it intentionally challenging for our scheduler to stay below \qty{1}{\percent} by shifting each benchmark's mean change for the full execution to \qty{3}{\percent}.
This is an intentionally hard regime for controlling the false-positive rate, as we defined everything above \qty{3}{\percent} as the edge at which we should claim performance changes.
We push to both $+$ and $-\qty{3}{\percent}$ and use the worse option for validating $\alpha$.
We replay the full measurement pool across all traces and run the full scheduler end to end on that data.
For our self-recorded Go corpus, we conduct direct A/A benchmarks, i.e., any detected change is automatically a false positive.
The scheduler early-stops under anytime-valid stopping rules.
Undecided benchmarks run the full measurements, and we claim each benchmark with the bootstrap CI detector.
After the full replay, we compare the reported changes against the ground truth and evaluate the observed $\alpha$ level.
The classical, mathematically guaranteed error control under confidence sequences breaks, since $\alpha$ becomes a union over two claiming mechanisms.

\paragraph{E2: Budget Saved at Matched Recall}
Experiment E2 evaluates how much execution time we save at held recall and false-positive rate compared to the best uniform baseline.
Both the uniform baseline and \sysname replay the same data such that we can attribute savings to the scheduling policy and early stopping.
Our system stops each benchmark when its rules fire or when a global budget cap is reached.
To find the best $N^\star$ uniform configuration, we measure the recall for a grid of different execution depths.
If, for instance, the grid reaches \qty{88}{\percent} recall at depth $80$ and \qty{90.5}{\percent} at the next depth, we interpolate between the two depths to match \sysname at a \qty{90}{\percent} recall level.
Interpolation only removes an upward bias in the baseline's budget; \sysname's own budget is measured exactly.
We evaluate recall and FP rate against the exhaustive run effects, such that we can trace disagreements to our early stopping.

\paragraph{E3: Detection Latency}
To isolate the effects of our decision-closest prioritization, we run the scheduler under the same data with our prioritization enabled compared to a round-robin baseline.
We measure at what execution time the scheduler surfaces regressions under both scheduling policies.
We compare at what budget we reach \qty{50}{\percent}, \qty{75}{\percent}, and \qty{90}{\percent} of the changes.
Our policy works by front-loading the benchmarks that show high magnitude effects early on.

\paragraph{E4: Scheduler Overhead}
For the scheduler overhead, we replay the recorded dataset.
The values are pre-drawn before starting the experiment; we measure the benchmark selection and update step as described in \cref{alg:scheduler}.
The benchmark invocations are therefore no-ops, i.e., we remove the benchmark execution time and focus the measurements on the overhead.

\subsection{Results}
\subsubsection{False-Positive Control}
\label{subsubsec:fpr-results}
State-of-the-art methods that optimize for measurement stability optimize costs but have limited control over statistical errors.
On the four Go suites, we recorded A/A benchmarks, i.e., all changes in that realm are false positives.
We have \num{254} benchmarks for the four Go projects with \num{127000} measurements and zero false positives.
The recording includes one A/A run and one A/B run for each project.
\cref{tab:govalid} summarizes the Go corpus and its per-project results; for the intentionally challenging regime that shifts each mean onto the \qty{3}{\percent} decision boundary, we report the false-positive rate at the suite level.
On the Go suite the worse edge sits at \qty{0.79}{\percent}.
The FP rate holds on every Java project for the A/A experiments (\cref{tab:perproject}) with the worst at $\qty{0.87}{\percent} < \qty{1}{\percent}$ for the eclipsecollections project.
This worst case rate decomposes into the confidence sequence error rate (\qty{0.46}{\percent}) and the bootstrap CI one (\qty{0.41}{\percent}).
The undecided benchmarks under the confidence sequence are decided by the bootstrap CI at full budget; this claims the remaining changes, and on the Go suites the full protocol reaches recall $1.000$ in the savings evaluation below.

\begin{table}[t]
  \centering
  \caption{Overview of the Go dataset, presenting early recall (recall from confidence-sequence stops alone), savings vs.\ the full run, and false positives. The suite-level A/A rate at the worse band edge (\qty{0.79}{\percent}) holds below the $\alpha=\qty{1}{\percent}$ target.}
  \label{tab:govalid}
  \scriptsize
  \setlength{\tabcolsep}{1.75pt}
  \begin{tabular}{llrrccr}
    \toprule
    suite & pair & \shortstack[c]{benchmarks\\A/A$+$A/B} & \shortstack[c]{changes\\(A/B)} & \shortstack[c]{early\\recall} & FP & saved \\
    \midrule
    prometheus/common  & v0.45.0,v0.46.0 & $34{+}34$ & $3$ & $1.00$ & $0$ & \qty{87.7}{\percent} \\
    klauspost/compress & v1.18.0,v1.18.2 & $18{+}18$ & $4$ & $.75$ & $0$ & \qty{83.4}{\percent} \\
    go-toml            & v2.1.1,v2.2.0   & $35{+}35$ & $8$ & $1.00$ & $0$ & \qty{86.5}{\percent} \\
    uber-go/zap        & v1.26.0,v1.27.0 & $40{+}40$ & $6$ & $.83$ & $0$ & \qty{83.2}{\percent} \\
    \bottomrule
  \end{tabular}
\end{table}

\subsubsection{Recall and Budget Savings at Matched Recall}
Our approach's main motivation is cost saving while keeping recall high and the false-positive rate low, i.e., we reach the same conclusions as the best uniform configuration more efficiently.
\sysname saves up to \qty{42}{\percent} of execution time against $N^\star$ at matched recall.
Across recall levels $0.7$--$0.9$ we save \qtyrange{29}{42}{\percent} of the $N^\star$ budget at a false-positive rate no worse than the baseline, so the savings cost neither recall nor precision.
At the $0.9$ recall level, this corresponds to detecting \qty{90}{\percent} of the performance changes at \qty{58}{\percent} of the execution time of the best uniform configuration.

The savings generalize across the Java corpus: seven of eight projects save \qtyrange{34}{61}{\percent} against $N^\star$ at matched recall as shown in \cref{tab:perproject}; jenetics tops out at recall $0.895$, so we compare it at $0.85$, saving \qty{27}{\percent}.
On Go, running \sysname{} to self-termination saves \qtyrange{83}{88}{\percent} of the full run per project (\cref{tab:govalid}), at recall $1.000$ with zero false positives.

Once we accumulate sufficient evidence, we immediately retire the benchmark instead of running it to a fixed depth.
The baseline keeps running already-decided benchmarks; we reclaim that wasted execution time without sacrificing recall.
When running \sysname until self-termination, meaning benchmarks either early-stop or replay the full measurements if undecided, we land at \qty{95.9}{\percent} recall with \qty{29.6}{\percent} saved against the full run.

Under self-termination the recall spans $0.90$--$0.97$ and the false-positive rate never exceeds $\alpha = \qty{1}{\percent}$.

On the Go data the same protocol saves \qtyrange{19}{83}{\percent} at recall $0.7$--$0.9$ and \qty{47.0}{\percent} at $0.95$ against $N^\star$, and self-terminates at \qty{85.4}{\percent} saved against the full run with recall $1.000$ at zero false positives.

\begin{figure}
  \centering
  \includegraphics[width=0.78\columnwidth]{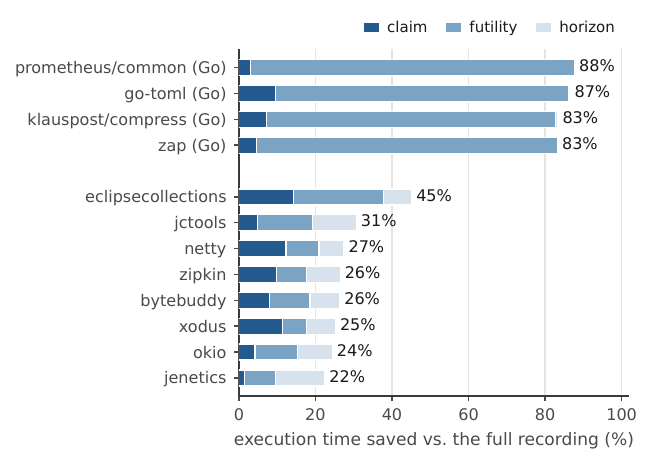}
  \caption{Contribution of each stopping rule (claim, futility, horizon) to the per-project savings. Bar totals are the self-termination savings vs.\ the full run (\qty{29.6}{\percent} aggregate on Java, \qty{85.4}{\percent} on Go), complementing the matched-recall savings vs.\ $N^\star$ in \cref{tab:perproject}.}
  \label{fig:savingsprojects}
\end{figure}

On the Java suite, the different stopping strategies contribute with \qty{29}{\percent}, \qty{42}{\percent}, and \qty{29}{\percent} to the overall savings, where no strategy overly dominates (\cref{fig:savingsprojects}).
The savings gained by each strategy depend on the nature of the suite.

When only considering regressions, \sysname performs even better.
Early stopping more aggressively and ignoring the positive changes, we save \qtyrange{56}{63}{\percent} against $N^\star$ across recall $0.7$--$0.9$ ($\alpha = \qty{1}{\percent}$) and \qty{55.5}{\percent} against the full run under self-termination; the one-sided recall reaches $0.926$.

\begin{table}[t]
  \centering
  \caption{Overview of the savings, reached recall, FP rate and A/A false-positive rate across the Java dataset.}
  \label{tab:perproject}
  \footnotesize
  \setlength{\tabcolsep}{4pt}
  \begin{tabular}{lrrrrr}
    \toprule
    project & benchmarks & saved vs.\ $N^\star$ & recall & FP & A/A \\
    \midrule
    bytebuddy          & $912$  & \qty{36}{\percent} & $0.97$ & $.000$ & $.005$ \\
    eclipsecollections & $1350$ & \qty{61}{\percent} & $0.97$ & $.002$ & $.009$ \\
    jctools            & $1055$ & \qty{44}{\percent} & $0.94$ & $.001$ & $.006$ \\
    jenetics           & $891$  & \qty{27}{\percent} & $0.90$ & $.000$ & $.009$ \\
    netty              & $1281$ & \qty{36}{\percent} & $0.97$ & $.005$ & $.007$ \\
    okio               & $1498$ & \qty{45}{\percent} & $0.96$ & $.000$ & $.005$ \\
    xodus              & $349$  & \qty{34}{\percent} & $0.96$ & $.009$ & $.007$ \\
    zipkin             & $1056$ & \qty{40}{\percent} & $0.95$ & $.002$ & $.006$ \\
    \bottomrule
  \end{tabular}
\end{table}

\subsubsection{Detection Latency}
Especially under low budgets, our prioritization policy shortens the time to detection for relevant performance changes.

Our prioritization detects a given share of change at \qtyrange{23}{58}{\percent} of the budget round-robin needs on the Java corpus (Table~\ref{tab:latency}).
For the Go data prioritization needs \qtyrange{17}{22}{\percent} of the round-robin budget.

Beyond detected changes, practitioners are especially interested in finding high magnitude changes first.
Therefore we report magnitude-weighted change detection:
We calculate the absolute mean change detected as a fraction of the total change magnitude.
Our decision-closest policy favors high-magnitude changes early on, widening the gap to round-robin even further.
We capture \qty{50}{\percent} of the total change magnitude at \qty{7.9}{\percent} of the budget while round-robin needs \qty{51.8}{\percent} on the Java corpus (\qty{1.6}{\percent} vs \qty{8.3}{\percent} on Go).

The effect is broad: we detect sooner on all $8$ of $8$ Java projects and all $4$ of $4$ Go ones.

\begin{table}[t]
  \centering
  \caption{We compare the change detection point as a fraction of the overall execution duration across Java and Go projects, highlighting that prioritization surfaces changes sooner.}
  \label{tab:latency}
  \footnotesize
  \setlength{\tabcolsep}{4pt}
  \begin{tabular}{lrrrr}
    \toprule
    & \multicolumn{2}{c}{changes detected} & \multicolumn{2}{c}{magnitude captured} \\
    \cmidrule(lr){2-3} \cmidrule(lr){4-5}
    & \sysname & r-robin & \sysname & r-robin \\
    \midrule
    \multicolumn{5}{l}{Java} \\
    \qty{50}{\percent} & \qty{14.9}{\percent} & \qty{64.4}{\percent} & \qty{7.9}{\percent}  & \qty{51.8}{\percent} \\
    \qty{75}{\percent} & \qty{29.6}{\percent} & \qty{75.1}{\percent} & \qty{20.4}{\percent} & \qty{73.0}{\percent} \\
    \qty{90}{\percent} & \qty{45.7}{\percent} & \qty{78.9}{\percent} & \qty{36.0}{\percent} & \qty{77.7}{\percent} \\
    \midrule
    \multicolumn{5}{l}{Go} \\
    \qty{50}{\percent} & \qty{2.0}{\percent}  & \qty{11.8}{\percent} & \qty{1.6}{\percent}  & \qty{8.3}{\percent}  \\
    \qty{75}{\percent} & \qty{2.6}{\percent}  & \qty{12.5}{\percent} & \qty{2.2}{\percent}  & \qty{12.2}{\percent} \\
    \qty{90}{\percent} & \qty{3.6}{\percent}  & \qty{16.4}{\percent} & \qty{2.8}{\percent}  & \qty{13.0}{\percent} \\
    \bottomrule
  \end{tabular}
\end{table}

\subsubsection{Scheduler Overhead}
\label{subsec:overhead}
To validate practical applicability, our scheduler must not add significant overhead; otherwise, it inflates the end-to-end experiment duration.
Optimizing execution duration itself with a computationally expensive protocol might not yield any meaningful improvement in practice.
On the Java suite the scheduler spends \qty{8.9}{\micro\second} per measurement across \num{8392} benchmarks and \num{7.3} million measurements, which accumulates to \qty{64.9}{\second} of scheduling, or \qty{0.00089}{\percent} of the \qty{2020}{\hour} of benchmark execution it schedules (\qty{0.24}{\second} of \qty{5.3}{\hour} on Go).

\section{Discussion}
\label{sec:discussion}
In the following, we discuss design trade-offs, when our approach pays off, and threats to validity.

\subsection{Design Trade-offs}
While our early stopping rules control false positives, they risk false negatives: we may retire benchmarks that would reveal regressions given more measurements.
This applies to the horizon stopping in particular.
Such a missed regression can be costly at operational scale.
We therefore report savings only at recall matched to the baseline, such that we do not sacrifice detection quality for execution time reduction.
Empirically, \sysname reaches the same recall regime ($0.90$--$0.97$) as the best uniform baseline across all eight Java projects, at a false-positive rate that never exceeds $\alpha = \qty{1}{\percent}$ (\cref{tab:perproject}).

\paragraph{Confidence Sequence Tax}
The confidence sequence intervals are wider than a fixed-$n$ confidence interval, as they must maintain the anytime-valid property.
Under an oracle that could determine the ideal fixed-$n$ duration a priori, \sysname would always spend more time, since we pay this tax.
However, a method to accurately approximate the oracle is practically infeasible, since a noisy neighbor can already invalidate the precomputed duration.
We minimize this tax with growth-rate-optimal betting, which optimizes evidence accumulation such that the confidence sequence converges faster.
In addition, for benchmarks that consume all the available execution time, we fall back to bootstrapped CIs to avoid the anytime-valid tax.
This is valid since the horizon is fixed a priori, and we share the false-positive rate across the sequential CS and the terminal bootstrap CI.

\subsection{When Adaptive Scheduling Pays Off}
Adaptive scheduling works particularly well on heterogeneous suites with easy and hard benchmarks: the easy ones resolve quickly, such that we can concentrate on the hard ones.
Accordingly, the per-project savings against $N^\star$ range from \qtyrange{27}{61}{\percent} (\cref{tab:perproject}), highlighting the different headroom across suites.
Beyond stopping, our prioritization surfaces the large changes that matter most in practice sooner~(\cref{tab:latency}).
In homogeneous suites, by contrast, uniform allocation is already near-optimal, provided a developer identifies the right fixed-$n$ execution time.

\subsection{Threats to Validity}
\sysname's savings rest on assumptions regarding statistical model, baseline, labels and replayed dataset, which we discuss below.

\paragraph{Internal Validity}
Because we split the false-positive rate across two mechanisms---the sequential test and the terminal fixed-$n$ bootstrap CI---we do not prove the combined error guarantee analytically, but instead validate it empirically.
The E1 experiment confirms the combined guarantee of $\alpha = \qty{1}{\percent}$ holds on every Java project and at suite level on Go, even under the hardest scenario.
The false-positive rate holds per benchmark; we do not control the suite level error.
We derive the performance-change labels by replaying all available measurements under resampling with a bootstrap CI detector.
Since we have no externally verified oracle, we evaluate against the conclusion of a full, exhaustive run instead.

\paragraph{Construct Validity}
We choose the uniform baseline through a grid search against the full-run labels.
Specifically, for each recall level, we select the $N^\star$ that reaches it with the least execution time.
In practice, a practitioner cannot know $N^\star$ and either overcommits, wasting budget, or undercommits, missing regressions.
We therefore compare against a generous baseline, which makes our reported savings conservative.
Benchmarks with true effects near our \qty{3}{\percent} decision boundary~\cite{japke2025muoptimestaticallyreducingexecution} are hard to decide and sensitive to sampling variation.

\paragraph{External Validity}
Anytime-valid confidence sequences require i.i.d.\ data, an assumption supported by RMIT measurements.
We therefore replay the corpus in the recorded RMIT order.
The results on our replayed Go corpus hold the false-positive rate under $\alpha$ while saving \qtyrange{83}{88}{\percent} of the full run (\cref{tab:govalid}).
In contrast, the Java corpus was recorded sequentially, and we approximate RMIT by bootstrap-resampling.

\paragraph{Replay versus Live Execution}
We replay all recorded measurements, so the savings come from the scheduling policy alone.
A live run additionally pays for JVM warmup, benchmark switching, and duet synchronization, so the replayed savings are an upper bound.

Beyond the recording order, the dataset has a heterogeneous distribution of measurement depths (median $1490$, minimum $60$ measurements per benchmark).
In particular, shallow benchmarks leave little headroom for cutting execution time.
Conversely, deployments with deeper recordings offer greater savings potential for \sysname.

\section{Conclusion}

\sysname schedules a microbenchmark suite under adaptive prioritization and early stopping to cut \qty{42}{\percent} of aggregate execution time on the Java suite, and up to \qty{61}{\percent} on individual projects, at matched false-positive rate and recall against $N^\star$; on Go, it saves \qty{85}{\percent} of the full run at full recall with zero false positives.
It prioritizes the benchmarks nearest a decision to surface regressions sooner and stops them once sufficient evidence has accumulated.
We model each benchmark as a betting process against the null hypothesis of no relevant change.
In the presence of a performance change, the bets accumulate wealth (evidence) over time.
Once we reach sufficient evidence, we stop the benchmark under measured false-positive control with no historical data or code features needed.
To validate our false-positive control, we shift each benchmark by a constant offset onto the decision boundary ($\pm \qty{3}{\percent}$ mean change).
Across the 8 Java projects and at suite level on the 4 Go projects, the false-positive rate never exceeds the target $\alpha$ of \qty{1}{\percent}.
As AI accelerates software change, \sysname makes continuous performance assurance more affordable.

As future work, the anytime-valid statistical machinery enables extensions to more prioritization strategies that consider feature inputs such as code changes or historical information.
Static offline optimizations such as \textmu OpTime~\cite{japke2025muoptimestaticallyreducingexecution}, for instance, can act as a complementary signal that feeds into the scheduling decision as a prior.
It is worthwhile to explore more sophisticated prioritization strategies, considering the computational complexity to keep the scheduler's overhead low.

\section*{Acknowledgments}
We used Claude Code (Anthropic) to assist with the implementation of the AdapBench prototype.
The authors made all design decisions, reviewed all generated code, and validated all experimental results.

\bibliographystyle{IEEEtranN}
\bibliography{bibliography}

\end{document}